\documentclass{cpp2e}
\usepackage{epsfig, floatflt, rotating, latexsym, amssymb}
\newcommand{\beq}{\begin{equation}}
\newcommand{\eeq}{\end{equation}}
\newcommand{\bea}{\begin{eqnarray}}
\newcommand{\eea}{\end{eqnarray}}

\newcommand{\gcc}{{\rm~g\,cm}^{-3}}
\newcommand{\ApJ}[1]{Astrophys.~J. {\bf #1}}
\newcommand{\AandA}[1]{Astron.\ Astrophys. {\bf #1}}

\newcommand{\PR}[2]{Phys.\ Rev.\ #1 {\bf #2}}

\shorttitle{Outer Envelopes of Neutron Stars}
\shortauthor{A.Y. Potekhin, G. Chabrier, D.G. Yakovlev}
\contvolum{41} 
\contyear{2001} 
\contnumber{2--3} 
\contpages{in press} 

\begin{document}
\setcounter{page}{1}
\makeheadings

\title{Coulomb Plasmas in Outer Envelopes of Neutron Stars}

\author{A.Y. Potekhin$^{\rm a}$, G. Chabrier$^{\rm b}$, 
D.G. Yakovlev$^{\rm a}$}
\address{$^{\rm a}$Ioffe Physical-Technical Institute,
     194021 St.\ Petersburg, Russia\\
$^{\rm b}$%
     Ecole Normale Sup\'erieure de Lyon,
       CRAL, 
     69364 Lyon Cedex 07, France\\
     e-mail: palex@astro.ioffe.rssi.ru}
\date{\today}
\maketitle

\begin{abstract}
Outer envelopes of neutron stars 
consist mostly of fully ionized, strongly coupled Coulomb plasmas
characterized by 
typical densities $\rho\sim10^4$--$10^{11}\gcc$ and temperatures 
$T\sim10^4$--$10^9$~K. Many neutron stars possess magnetic fields 
$B\sim10^{11}$--$10^{14}$~G. 
Recent theoretical advances allow one to calculate
 thermodynamic functions and electron transport coefficients
for such plasmas with an accuracy required 
for theoretical interpretation of observations.
\end{abstract}

\section{Overview}
Envelopes of neutron stars (NSs) are divided in the inner and outer 
envelopes.
Properties of the inner envelopes
are outlined in a companion paper \cite{Yak}.
Here we focus on the outer ones,
at typical densities
$\rho\sim10^4-10^{11}\gcc$ and temperatures $T\sim10^4-10^9$~K.
These envelopes are relatively thin, but they
 affect significantly NS evolution.
In particular, they provide thermal insulation of stellar interiors.

Bulk properties of the NS envelopes (pressure $P$,
internal energy $U$) are determined mainly by degenerate electron gas.
The electrons are relativistic at $x_{\rm r}\gtrsim1$,
where 
$
    x_{\rm r}  \equiv  {\hbar k_{\rm F} / m c} \approx
       (\rho_6  Z / A)^{1/3}
$,
$k_{\rm F}=(3\pi^2n_{\rm e})^{1/3}$ is the Fermi wave number, $m$ is the electron mass,
$n_{\rm e}=Zn_{\rm i}$ is the electron number density,
$\rho_6\equiv\rho/10^6\gcc$,
$Z$ and $A$ are the ion charge and 
mass numbers, respectively.
Degeneracy is strong
at $T\ll T_{\rm F}=(\epsilon_{\rm F}-mc^2)/k_{\rm B}$, 
where $\epsilon_{\rm F}=mc^2\gamma_{\rm r}$
and  $\gamma_{\rm r}=\sqrt{1+x_{\rm r}^2}$ is the Lorentz factor.

The ionic component of the plasma can
form either strongly coupled Coulomb liquid or solid, classical 
or quantum. The classical ion coupling parameter is $\Gamma=(Ze)^2/ak_{\rm B}T$,
where 
$a$ is the ion-sphere radius
($\frac{4\pi}{3}a^3=n_{\rm i}^{-1}$).
The ions form a 
crystal if $\Gamma$ exceeds some critical
value $\Gamma_m$ (see below). 
The quantization of ionic motion becomes important
at $T\ll T_{\rm p}$, where $T_{\rm p}=\hbar 
\,\sqrt{4 \pi n_{\rm i} Z^2 e^2/ M}
   \,/k_{\rm B}$ is the ion plasma temperature,
and $M$ is the ion mass. 
The ionic contribution determines specific heat $C_V$, 
unless $T\ll T_{\rm p}$.

The ions and electrons are coupled together through the electron response
(screening). The inverse screening (Thomas--Fermi) length is 
$
   k_{\rm TF} = ( 4\pi e^2\, \partial n_{\rm e} / \partial \mu)^{1/2},
$
where $\mu$ is the electron chemical potential. If $T\ll T_{\rm F}$, then
$\mu\approx\epsilon_{\rm F}$ 
and $k_{\rm TF}^2/k_{\rm F}^2\approx4(\alpha/\pi)
\,\gamma_{\rm r}/x_{\rm r}$,
where $\alpha$ is the fine structure constant.
Note that $\gamma_{\rm r}/x_{\rm r}\to1$ at $x_{\rm r}\gg1$;
therefore the screening effects do not vanish
even at high densities.

Thermodynamic properties are altered by the magnetic field $B$ in the case
where the Landau quantization
of transverse electron motion is important.
The field is called {\em strongly quantizing\/} when it sets 
all electrons on the ground Landau level.
This occurs at $\rho<\rho_B$ and $T\ll T_B$ (see, e.g., 
ref.~\cite{YaK}),
where
$
   \rho_B\approx 7045\,(A/Z)\,B_{12}^{3/2}\gcc,
   T_B\approx 1.343\times10^8(B_{12}/\gamma_{\rm r}){\rm~K},
$
and $B_{12}\equiv B/10^{12}$~G.
{\em Weakly quantizing\/} fields ($\rho_B<\rho$) do not significantly
alter $P$ and $U$ but cause oscillations
of $C_V$, other second-order quantities, and
electron transport coefficients with increasing density.

Characteristic domains in the $\rho$--$T$ diagram are shown in 
Figure~\ref{fig-domains}.
The short-dashed lines on the left panel
indicate the region of partial ionization.
The dot-dashed lines correspond to 
$\Gamma=1$ (upper lines)
and $\Gamma=175$ (liquid/solid phase transition).
Long-dashed lines on
the right panel separate three $\rho$--$T$ regions
where the magnetic field is
strongly quantizing
(to the left of $\rho_B$ and considerably below $T_B$),
weakly quantizing
(to the right of $\rho_B$ at $T\lesssim T_B$),
or classical (much above $T_B$).
          \begin{floatingfigure}{10.2cm}
            \mbox{\epsfig{file=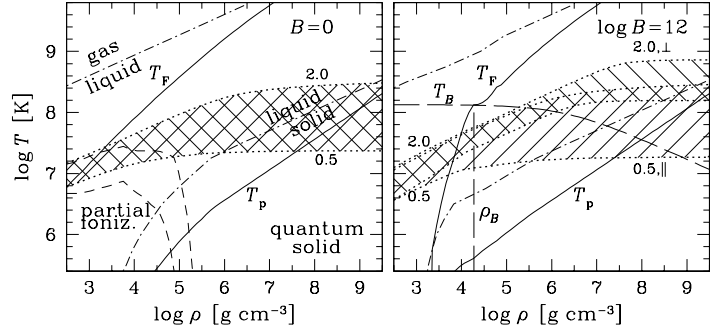, width=9.5cm}}
     \caption{%
Characteristic $\rho$--$T$ domains for 
outer NS envelopes composed of iron.
Left panel: non-magnetic plasma; right panel: $B=10^{12}$~G.
Shown are $\rho$-dependences of $T_{\rm F}$, $T_{\rm p}$,
$T_B$, $\rho_B$ (see text);
lines where $\Gamma=1$ and $\Gamma=175$
(dot-dashed curves);
lines where effective $Z$ equals 15 or 20 
(short-dashed lines on the left panel).
Dotted lines and hatched regions show temperature profiles
in NSs at different cooling stages
 and (on the right panel) at different inclinations
 of the magnetic field.
}
\label{fig-domains}
       \end{floatingfigure}
%
The dotted lines in Fig.~\ref{fig-domains}
show profiles $T(\rho)$
in the envelope of a ``canonical'' NS with mass 
$1.4\,M_\odot$ and radius 10 km, and
with an effective surface temperature $T_{\rm s}=5\times10^5$~K
or $2\times10^6$~K
(the values of $T_{\rm s6}\equiv T_{\rm s}/10^6{\rm~K}$
are marked near these curves).
Typical temperatures of isolated 
NSs are believed to lie 
in the hatched region between these two lines.
In a magnetized NS, $T(\rho)$ depends
on strength as well as direction of the magnetic field. 
Therefore on the right panel we show {\em two\/} dotted 
curves for each value of $T_{\rm s6}$:
the lower curve of each pair corresponds to the heat propagation
along the field lines ($\|$, -- i.e., near the magnetic poles)
and the upper one to the transverse propagation ($\perp$,
near the magnetic equator).

\section{Equation of state}
Thermodynamic functions of the electron gas at arbitrary degeneracy
are expressed through the well known Fermi-Dirac integrals.
For astrophysical use, it is convenient to employ analytic fitting
formulae for these functions presented, e.g., in ref.~\cite{CP98}.

Nonideal (exchange and correlation) corrections for nonrelativistic
electrons at finite temperature have been calculated and parameterized 
in ref.~\cite{IIT}. For the relativistic electrons at low $T$,
an analytic expansion of the exchange corrections is given, e.g., 
in ref.~\cite{SB96} (in this case, the correlation corrections are negligible).
A smooth interpolation between these two cases has been constructed
in ref.~\cite{PC00}.

For the ionic component at $\Gamma\gg1$, 
the main contribution to the thermodynamic
functions comes from the ion correlations.
Strongly coupled one-component Coulomb plasmas (OCP) 
of ions in the {\em uniform\/}
electron background have been studied by many authors.
The thermodynamic functions of
the classical OCP liquid have been calculated 
\cite{DWSC,Caillol}
at $\Gamma\geq1$
and parameterized \cite{CP98,PC00}
for $0 < \Gamma\lesssim200$. 
The latter parameterization
ensures accuracy $\sim10^{-3}k_{\rm B} T$ (per particle).

For the classical Coulomb crystal, accurate numerical results
and fitting formulae to the free energy $F$ (with anharmonic
corrections taken into account) have been presented 
in refs.~\cite{Dubin,FH}. A comparison of
the latter results with the fit \cite{PC00} for the liquid
yields the liquid--solid phase transition 
at $\Gamma_m=175.0\pm0.4$ (curiously, this value 
could be derived in ref.~\cite{DWSC},
but it was first noticed in ref.~\cite{PC00}).

In the solid phase, 
quantum effects may be important
at the considered temperatures. 
These effects can be most easily taken into account in
the approximation of a harmonic Coulomb crystal.
A convenient analytic approximation 
is provided by a model \cite{C93} 
which treats two phonon modes as Debye modes and the third one
 as an Einstein mode.
Numerical calculations \cite{Baiko} of the harmonic-lattice
contributions to $F$, $U$, $C_V$,
and the mean-square ion displacement 
are reproduced by this model at arbitrary $T/T_{\rm p}$
within several percent, 
which is sufficient for many applications.

\begin{floatingfigure}{8cm}
            \mbox{\epsfig{file=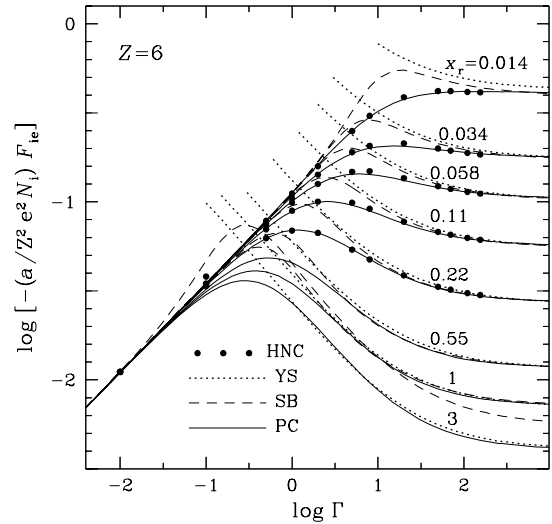, width=7.3cm}}
     \caption{%
Nonideal $ie$
contribution to the free energy of carbon
(normalized to $N_{\rm i} (Ze)^2/a$, where $N_{\rm i}$ is the number of ions)
at 8 values of 
$x_{\rm r}$ marked near the curves.
HNC results \protect\cite{CP98} 
and relativistic perturbation theory
\protect\cite{YS} (YS) are compared with
analytic fits \protect\cite{SB} (SB) and \protect\cite{PC00} (PC).
}
\label{fig-ie}
       \end{floatingfigure}
Finally, there is a contribution 
from ion-electron ($ie$) correlations,
which can be treated as polarization of the ``jellium" of the degenerate
electrons. In a Coulomb liquid, this contribution has been evaluated
in refs.~\cite{GalamHansen,YS} using a perturbation theory
which involves the electron dielectric function
and the static structure factor of ions.
This approximation is justified at $k_{\rm TF}\ll k_{\rm F}$
(which is the case at $\rho_6\gtrsim10^{-2}$).
At smaller densities the $ie$ contribution 
in the Coulomb liquid has been calculated in
ref.~\cite{CP98} using the HNC technique
(heavy dots in Fig.~\ref{fig-ie}).

In a Coulomb solid, the $ie$ contribution
 has been evaluated in ref.~\cite{PC00}
using the perturbation theory \cite{GalamHansen,YS}
and a model lattice structure factor. 
These results, though approximate, indicate that
the polarization corrections
may be rather important.
In particular, $\Gamma_m$ may be shifted by $\sim10$\%
when the $ie$ corrections in the liquid and solid phases
are taken into account.

The numerical results have been fitted \cite{CP98,PC00}
by analytic functions of $x_{\rm r}$ and $\Gamma$.
The approximation \cite{PC00} for the OCP liquid is shown
in Fig.~\ref{fig-ie} by solid lines.
Since the $ie$ correction is not dominant,
the accuracy of the fitting formulae ($\lesssim10$\%) 
is sufficient for most of astrophysical applications.
An alternative Pad\'e approximation,
proposed in ref.~\cite{SB} for the electron-ion fluid,
is also accurate at $x_{\rm r}\lesssim1$ for small and 
large $\Gamma$, but it is less accurate at intermediate
$\Gamma$ and inapplicable at $x_{\rm r}>1$ (dashed lines).

As mentioned above, quantizing magnetic fields significantly affect
the equation of state. In the fully ionized dense plasma,
the main effects are described by the model of ideal electron gas,
studied by many authors 
(e.g., \cite{YaK,BH}).

\section{Electron conductivities}
Heat conduction in NS envelopes is provided mainly
by the electrons. The thermal conductivity tensor $\kappa$
is determined mainly by the electron-ion scattering. 
Calculation of $\kappa$ should take into account specific
features of the Coulomb plasmas in NS envelopes, 
quite different from the terrestrial liquid and solid metals: 
(i) in the liquid
phase, an incipient long-range order emerges at $\Gamma\gtrsim10^2$;
(ii) in the solid phase, there are usually many
Brillouin zones within the Fermi surface; (iii) in the solid phase, 
the approximation of one-phonon scattering fails
near the melting, which necessitates the use of a more
general expression for the structure factor of ions.
These features have been taken into
account in ref.~\cite{PBHY}, where analytic approximations 
to the electron transport coefficients have also been derived.

Magnetic field 
hampers electron transport across the field lines:
transverse thermal and electrical 
conductivities are reduced 
by orders of magnitude in typical NS envelopes.
Strongly quantizing field significantly changes 
also longitudinal conductivities;
weakly quantizing field
causes de Haas--van Alphen oscillations.
These effects have been outlined, e.g., in \cite{YaK}.
A unified treatment of the electron transport coefficients
in the domains of classical, weakly quantizing, and strongly quantizing
magnetic field
has been developed in \cite{P99}. 
A Fortran code which implements this treatment
 is available at 
 \verb|http://www.ioffe.rssi.ru/astro/conduct/|.
Using this code and solving the thermal diffusion equation,
we have calculated 
the $T(\rho)$ profiles shown in Fig.~\ref{fig-domains}. 

\begin{acknowledgements}
AYP is grateful to the theoretical astrophysics group
at Ecole Normale Sup\'erieure de Lyon 
for generous hospitality and financial support. 
The work of AYP and DGY has been supported in part by 
INTAS (grant 96-542)
and RFBR (grant 99-02-18099). AYP and DGY thank
 the Organizing Committee 
 for support provided to attend PNP10.
\end{acknowledgements}

\begin{received}
Received 29 September 2000
\end{received}

\end{document}